\documentclass[11pt]{article}
\usepackage[dvips]{graphicx}
\usepackage{latexsym}
\usepackage[latin1]{inputenc}

\def\Comment#1{}
\newcommand{\bean}{\begin{eqnarray*}}
\newcommand{\eean}{\end{eqnarray*}}
\newcommand{\gapproxeq}{\lower
.7ex\hbox{$\;\stackrel{\textstyle >}{\sim}\;$}}
\newcommand{\lapproxeq}{\lower
.7ex\hbox{$\;\stackrel{\textstyle <}{\sim}\;$}}

\newcommand\lsim{\mathrel{\rlap{\lower4pt\hbox{\hskip1pt$\sim$}}
    \raise1pt\hbox{$<$}}}
\newcommand\gsim{\mathrel{\rlap{\lower4pt\hbox{\hskip1pt$\sim$}}
    \raise1pt\hbox{$>$}}}
\newcommand{\ba}{\begin{array}}
\newcommand{\ea}{\end{array}}
\newcommand{\nn}{\nonumber}

\newcommand{\be}{\begin{equation}}
\newcommand{\ee}{\end{equation}}
\newcommand{\bear}{\begin{eqnarray}}
\newcommand{\eear}{\end{eqnarray}}

\newcommand{\cO}{{\cal O}}

\newcommand{\Frac}[2]{\frac{\displaystyle #1}{\displaystyle #2}}
\newcommand{\Int}{\displaystyle{\int}}

\begin{document}
\thispagestyle{empty}
\begin{center}
\hspace*{10cm} {\bf  } \\
\hspace*{10cm} {\bf   }\\
\vspace*{2cm}
\begin{Large}
{\bf Some Remarks on the Pad\'e Unitarization of\\ Low-Energy Amplitudes
}
 \\[1cm]
\end{Large}
{ \sc P.~Masjuan},$^a$ { \sc J.J. Sanz-Cillero}$\,^a$ and { \sc J.~Virto}$\,^b$  \\[0.5cm]

{\footnotesize
\emph{$^a$ IFAE, Universitat Autònoma de Barcelona, 08193 Bellaterra, Barcelona, Spain}\\
\emph{$^b$ Università di Roma ``La Sapienza'' and INFN Sezione di Roma, 00185 Roma, Italy}\\
}

\vspace{0.5cm}

{\footnotesize May 21, 2008}

\vspace{0.5cm}
\begin{abstract}
We present a critical analysis of Padé-based methods for the unitarization of low energy amplitudes. We show that the use of certain Pad\'e
Approximants to describe the resonance region may lead to inaccurate determinations. In particular, we find that in the Linear Sigma Model the
unitarization of the low energy amplitude through the inverse amplitude method produces essentially incorrect results for the mass and width of the
sigma. Alternative sequences of Padés are studied and we find that the diagonal sequences (i.e., $[N/N]$) have much better convergence properties.
\end{abstract}
\end{center}




%
%
%
\vspace{0.1cm}

\section{Introduction}

\quad Effective Field Theories (EFT) have become  a very useful tool for the description of low-energy physics~\cite{EFT}. Based on symmetry and
dimensional analysis arguments, they allow to organize a perturbative expansion of the amplitudes in powers of soft momenta over some characteristic
scale, $p/\Lambda$. For instance, below the lightest resonance multiplet, the interactions between the pseudo-Nambu-Goldstone bosons from the
spontaneous chiral symmetry breaking -the pions in the $SU(2)$ case- is provided by Chiral Perturbation Theory
($\chi$PT)~\cite{ChPT-Weinberg,oneloop,ChPT-SU3}. Its range of validity is a few hundred MeV, breaking down as one approaches new states not included
in the EFT. Thus, $\chi$PT is unable to describe the resonance region and one needs to incorporate additional ingredients.

Several works have tried to extend  the range of validity of the EFT by means of the unitarization of the low-energy amplitude. Unitarization methods
have been applied extensively to Quantum Chromodynamics and $\chi$PT~\cite{unitariza}, where the issue of the scalar resonances is particularly
interesting. However, their range of applicability is wider; for example they have also been applied to $WW$--scattering~\cite{BSM}. Pad\'e
Approximants (PAs)~\cite{pipi-Pade,remarks-unitariza} and the Inverse Amplitude Method (IAM)~\cite{IAM,IAM-op6} are among the most usual ones,
although some remarks and criticisms on the reliability of these and other unitarization methods have been
raised~\cite{remarks-unitariza,Leutwyler-criticism,IAM-P12-critics}. In the present letter we focus our attention on the PAs. We study up to what
point one can rely on them to describe the resonant properties of the theory. By means of a couple of models (the Linear Sigma Model and a vector
resonance model),  we show that these unitarization procedures may lead to improper determinations of the resonance pole position (masses and
widths)~\cite{IAM-poles}. Furthermore, one may not recover the right values for the Low-Energy Couplings (LECs) of the EFT if the PAs are applied to
describe the resonant region.

The starting point of our analysis is, therefore, a model where the properties of the resonances are known. Then we derive and unitarize the
corresponding low-energy amplitude. The predictions for the masses and widths obtained from the PA  sequence $[1/N]$ (referred to as IAM in some
works~\cite{remarks-unitariza,IAM,IAM-op6,IAM-P12-critics}) are compared and found to be quite different from those of the original model. As an
alternative to the badly behaved sequence $[1/N]$, we propose the use of the PA sequence $[N/N]$, which quickly converges as $N$ is increased. For
simplicity the chiral limit will be assumed all along the paper, but this will not alter the main conclusion.

The structure of this letter is the following. We begin, in Section
\ref{LSM}, with the analysis in the LSM: in Section \ref{LSM1} we
compute the exact position of the pole of the sigma correlator at
the one-loop level. Then in Section \ref{LSM2} we consider the Linear Sigma Model (LSM)
at low energies and present the $\pi\pi$ scattering amplitudes. In
Section \ref{LSM3} we apply the IAM procedure to the partial wave
amplitudes to recover the mass and width of the sigma. In Section
\ref{padessec} we analyze PA sequences at higher orders. In Section
\ref{VM} the corresponding results are discussed in the context of a
vector resonance model. We then conclude, in Section \ref{conc},
with a short discussion.

\section{One-loop Linear Sigma Model} \label{LSM}

\subsection{Sigma pole position up to $\cO(g)$} \label{LSM1}

\quad We begin presenting the results for the one-loop corrections
to the mass and the width of the sigma in the LSM. At tree-level,
the sigma mass is
found to be $M_\sigma^2=2 \mu^2$, and the width is zero. At
next-to-leading order, the sigma pole gets shifted due to the
quartic potential, i.e., $M_\sigma^2= 2\mu^2 + \cO(g)$, and the
width becomes different from zero.

In order to determine the scalar meson mass and width up to $\cO(g)$, we compute the one-loop sigma correlator~\cite{oneloop},
\begin{equation}
i\Delta(s)^{-1}\, =\, s\, -\, M_\sigma^{2} \, \left[1 + \Frac{3 g}{16\pi^2}\, \left( -\Frac{13}{3} +  \ln\Frac{-s}{M_\sigma^2} + 3 \rho(s) \ln{\left(
\Frac{\rho(s)+1}{\rho(s)-1}\right)}  \right) + \cO(g^2) \right]\, ,
\end{equation}
where $\rho(s)\equiv \sqrt{1- 4 M_\sigma^2/s}\ $  and the term  $-13/3$
is determined by the renormalization scheme chosen by Ref.~\cite{oneloop},
which sets the relation $2 g F^2=M_\sigma^2$ at the one-loop order, with $F$ the
pion decay constant. Now it is
possible to extract the pole $s_p$ of the propagator up to the considered order in perturbation theory. If one approaches the branch cut from the
upper part of the complex $s$--plane, the pole in the second Riemann sheet is located at
\begin{eqnarray}
s_p &=&
\, M_\sigma^{  2} \,\left[\, 1\, +   \Frac{3 g}{16\pi^2}\,
\left( -\Frac{13}{3} +  \pi\sqrt{3}  \,  - \,  i \pi  \right) + \cO(g^2) \right]  \, ,
\end{eqnarray}
where we have used $s_p=M_\sigma^2 +\cO(g)$. The pole mass and width, defined from $s_p=(M_p-i \Gamma_p/2)^2$, are then given by
\begin{eqnarray}
\Frac{M_p^2}{M_\sigma^{  2} } &=& 1 \, +   \Frac{3 g}{16\pi^2}\, \left( -\Frac{13}{3} +  \pi\sqrt{3}  \right)+ \cO(g^2)\, \nonumber \\[2mm]
\Frac{M_p \Gamma_p}{M_\sigma^{  2}} &=&  \Frac{3 g }{16\pi} \,+\, \cO(g^2) \, .
\label{eq.LSM-pole}
\end{eqnarray}
As expected, at lowest order the pole width agrees with that derived from the decay amplitude~\cite{PW}.

\subsection{Low-energy expansion} \label{LSM2}

\quad We now consider the LSM at low energies. The contribution from
the sigma exchanges to the renormalized $\cO(p^4)$ $\chi$PT
couplings gives~\cite{oneloop}
\begin{eqnarray}
\ell_1^r(\mu) &=& \Frac{1}{4 g } + \Frac{1}{96\pi^2}\left[ \ln\Frac{M_\sigma^2}{\mu^2} -\Frac{35}{6} \right]  \, +\, \cO(g) \, , \nonumber \\[2mm]
\ell_2^r(\mu) &=& \Frac{1}{48\pi^2}\left[ \ln\Frac{M_\sigma^2}{\mu^2}
-\Frac{11}{6} \right]  \, +\, \cO(g) \, .
\label{eq.LSM-LECs}
\end{eqnarray}
The $\pi\pi$--scattering is determined by the $\pi^+\pi^- \to \pi^0\pi^0$ amplitude,
which is given up to $\cO(p^4)$ in the chiral expansion  by
\begin{eqnarray}
A(s,t,u) &=& \Frac{s}{F^2} \, +\, \Frac{2 s^2}{F^4} \ell_1^r \, +\, \Frac{s^2 + (t-u)^2}{2 F^4} \ell_2^r \\[2mm]
&& + \Frac{1}{96\pi^2 F^4} \left[ -3 s^2 \ln\Frac{-s}{\mu^2} - t(t-u) \ln\Frac{-t}{\mu^2} -u(u-t) \ln\Frac{-u}{\mu^2} + \Frac{5 s^2}{2} +\Frac{ 7
(t-u)^2}{6} \right]\, ,\nonumber
\end{eqnarray}
where $\mu$ refers here to the arbitrary renormalization scale,
and the chiral limit has been considered.

With this one constructs the definite isospin amplitudes
\begin{eqnarray}
T(s,t,u)^{\rm I=0} &=&
3 A(s,t,u) + A(t,s,u) +A(u,t,s) \, ,
\nn \\
T(s,t,u)^{\rm I=1} &=&
A(t,s,u) - A(u,t,s)\, ,
\nn \\
T(s,t,u)^{\rm I=2} &=&
A(t,s,u)+A(u,t,s) \, .
\end{eqnarray}
The partial wave projection is then provided by
\begin{equation}
t^I_J(s) \,\, =\,\, \Frac{1}{64\pi} \Int_{-1}^1 d\cos{\theta}
\, \, P_J(\cos{\theta})\,\,  T(s,t,u)^{\rm I}\, ,
\label{eq.PW}
\end{equation}
where $\theta$ is the scattering angle in the $\pi\pi$ center-of-mass rest frame.

Hence, for the first partial waves $t^I_J(s)$, with $IJ=00,11,20$, one finds the following $\cO(p^2)$ amplitudes,
\begin{eqnarray}
t_0^0(s)_{(2)} &=& \Frac{s}{16\pi F^2} \, , \nn \\[2mm]
t_1^1(s)_{(2)} &=& \Frac{s}{96\pi F^2}\, , \nn \\[2mm]
t_0^2(s)_{(2)} &=& -\Frac{s}{32\pi F^2}\, ,
\label{eq.t2}
\end{eqnarray}
and at $\cO(p^4)$,
\begin{eqnarray}
t_0^0(s)_{(4)} &=& \Frac{s^2}{48\pi F^4} \left[  11 \ell_1^r + 7 \ell_2^r  - \Frac{1}{96\pi^2} \left( 18\ln\Frac{-s}{\mu^2} + 7 \ln\Frac{s}{\mu^2}
-\Frac{51}{2}\right)\right] \, , \nn \\[2mm]
t_1^1(s)_{(4)} &=&\Frac{s^2}{96 \pi F^4} \left[  \ell_2^r -2  \ell_1^r  - \Frac{1}{96\pi^2} \left( \ln\Frac{-s}{\mu^2} - \ln\Frac{s}{\mu^2}
-\Frac{2}{3}\right)\right] \, , \nn \\[2mm]
t_0^2(s)_{(4)} &=& \Frac{s^2}{24\pi F^4} \left[  \ell_1^r + 2 \ell_2^r  - \Frac{1}{96\pi^2} \left( \Frac{9}{4} \ln\Frac{-s}{\mu^2} + \Frac{11}{4}
\ln\Frac{s}{\mu^2} -\Frac{51}{8}\right)\right] \, .
\end{eqnarray}
For the next section, it will be suitable to rewrite the $\cO(p^4)$ amplitudes in terms of the LSM parameters and the $\cO(p^2)$ scattering:
\begin{eqnarray}
t_0^0(s)_{(4)}\, &=& \, t_0^0(s)_{(2)} \, \, \times\,\, \Frac{11 s}{6 M_\sigma^{  2}} \left[  1  - \Frac{g}{264\pi^2} \left(
18\ln\Frac{-s}{M_\sigma^{  2}} + 7 \ln\Frac{s}{M_\sigma^{  2}} +\Frac{193}{3}\right) +\cO(g^2) \right] \, , \nn \\[2mm]
t_1^1(s)_{(4)} \, &=&\, t_1^1(s)_{(2)} \,\, \times\,\, \left(\Frac{-s}{M_\sigma^{  2}}\right)\,\, \left[1  + \Frac{g}{48\pi^2} \left(
\ln\Frac{-s}{M_\sigma^{  2}} - \ln\Frac{s}{M_\sigma^{ 2}} -\Frac{26}{3}\right) +\cO(g^2) \right] \, , \nn \\[2mm]
t_0^2(s)_{(4)} &=& t_0^2(s)_{(2)} \,\,\times\,\, \left( \Frac{- 2 s}{3 M_\sigma^{  2} } \right)\,\, \left[  1  - \Frac{g}{24\pi^2} \left(
\Frac{9}{4} \ln\Frac{-s}{M_\sigma^{  2}} + \Frac{11}{4} \ln\Frac{s}{M_\sigma^{  2}} +\Frac{163}{24}\right)+\cO(g^2) \right] \, ,
\label{eq.t4}
\end{eqnarray}
where we have used Eq.~(\ref{eq.LSM-LECs}) and
the relation $2 g F^2 =M_\sigma^2$~\cite{oneloop}.


\subsection{Unitarization of the $\chi$PT amplitude} \label{LSM3}

\quad The Inverse Amplitude Method (IAM
) provides an amplitude that is unitary not only at the perturbative
level but exactly. This means that in the elastic limit  one has for $s>0$
the partial-wave relation
\begin{equation}
\mbox{Im}\,  t(s)  \, =\, |t(s)|^2  \, ,
\end{equation}
where the indices $IJ$ are assumed ($t=t^I_J$). This relation can be reexpressed as a relation for the inverse amplitude:
\begin{equation}
\mbox{Im} \, t(s)^{-1}  \, =\, - 1  \, .
\end{equation}
Thus, the imaginary part of $t(s)^{-1}$ becomes completely determined
and one only needs to specify the real part Re$\,t^{-1}$. The IAM relies then on
a low-energy matching to $\chi$PT   (with $t_{_{\rm \chi PT}}^{-1}= t_{(2)}^{-1}\left[  1 - t_{(4)}/t_{(2)}
+...\right]$) in order to fix the unknown part of the amplitude.
Thus, at $\cO(p^4)$, one has the unitarized amplitude,
\begin{equation}
t_{_{\rm IAM}}\, \, =\, \, \Frac{t_{(2)}}{1\, -\, \Frac{t_{(4)}}{t_{(2)}}}\, .
\end{equation}
This expression is sometimes also known as a $P_1^1$ Pad\'e
Approximant  of the partial-wave amplitude.
The IAM has been also
extended up to $\cO(p^6)$ by means of what is sometimes named as a
$P^1_2$ approximant~\cite{IAM-op6,IAM-P12-critics}:
$$
t_{_{\rm IAM}}\, \, =\, \, \Frac{t_{(2)}}{1\, -\, \Frac{t_{(4)}}{t_{(2)}}
\, - \,  \Frac{t_{(6)}}{t_{(2)}}  + \left(\Frac{t_{(4)}}{t_{(2)}}\right)^2   }\, .
$$
%
%
%

However,
we want to remark that  $t(s)_{_{\rm IAM}}$ is not a PA in the variable $s$:
It is not a rational approximant since it also contains the logarithms from the
pion loops.
Thus, strictly speaking no theoretical argument ensures the recovery of
the physical amplitude.
Only in the tree-level limit $t(s)_{_{\rm IAM}}$  becomes a PA.
In any case, we will see that both the whole and the tree-level IAM amplitudes
are unable to reproduce the original partial waves in the resonance region.

Given the $\cO(p^2)$ and $\cO(p^4)$ $\chi$PT amplitudes from Eqs.~(\ref{eq.t2})--(\ref{eq.t4}), it is then possible to extract the poles of the
corresponding $t(s)_{_{\rm IAM}}$ for the LSM, satisfying $1= t(s)_{(4)}/t(s)_{(2)}$ at $s=s_p$:
\\
\\
{\bf IJ=00}
\begin{eqnarray}
s_p&=&   \Frac{ 6}{11}  M_\sigma^{  2} \, \left[  1  +  \Frac{g}{264\pi^2}
\left( \Frac{193}{3} + 25 \ln\Frac{6}{11} - 18 i \pi \right)
+\cO(g^2) \right] \, \, ,
\label{eq.LSM00}
\end{eqnarray}
{\bf IJ=11}
\begin{eqnarray}
s_p &=&  - M_\sigma^{  2} \,\,
\left[1  +  \Frac{g}{48\pi^2}
\left(   \Frac{26}{3} + i\pi \right)
+\cO(g^2) \right] \, ,
\label{eq.LSM11}
\end{eqnarray}
{\bf IJ=20 }
\begin{eqnarray}
s_p &=&    -\Frac{3}{2} M_\sigma^{  2} \,\,
\left[  1  + \Frac{g}{24\pi^2}
\left( \Frac{163}{24}+5 \ln\Frac{3}{2}+ \Frac{11 i \pi}{4}\right)+\cO(g^2) \right]
\, .
\label{eq.LSM20}
\end{eqnarray}
These are the poles that appear in the unphysical Riemann sheet as one approaches from upper half of the first Riemann sheet. There is also a
conjugate pole at $s_p^*$ if one approaches the real $s$--axis from below.

The first thing to be noticed is that poles appear in the $IJ=11$ and $20$
channels even for small values of $g $, contrary to what one expects in
the LSM, where no meson  with these quantum numbers exists. Furthermore,
these ``states'' are not resonances, as they are located on the left-hand
side of the complex $s$--plane, out of the physical Riemann sheet,
and carrying a negative squared mass.

As for the  $IJ=00$ channel, one finds a resonance with pole mass and width,
\begin{eqnarray}
\Frac{M_p^2}{M_\sigma^2}&=&  \Frac{6}{11} \, \, \left[  1  +  \Frac{g}{16\pi^2} \left( \Frac{50}{33} \ln\Frac{6}{11} +\Frac{386}{99}\right)
+\cO(g^2) \right] \, \, , \nn \\[2mm]
\Frac{M_p \Gamma_p}{M_\sigma^2}  &=& \Frac{24}{121}\,\cdot \,  \Frac{ 3 g}{16\pi}\,\, +\, \cO(g^2)  \, .
\end{eqnarray}
The IAM predictions for $M_p^2$ and $M_p \Gamma_p$ result,
respectively, 40\% and 80\% smaller than the original
ones in the LSM, computed in Eq.~(\ref{eq.LSM-pole}).
This points out the low reliability of this particular  method in order to recover
the hadronic properties of the theory from its effective
low-energy description.

The IAM poles remain badly located even in the weakly interacting limit, so this failure cannot be attributed to non-perturbative effects. In the
limit when $g \to 0$ and $M_\sigma $ is kept fixed one finds that the poles predicted in all the different channels fall down to the real $s$--axis.
We are left with just tree-level amplitudes and the expressions become greatly simplified. Due to the smoothness of this limit, it will be assumed in
the next analysis of higher order Pad\'e Approximants $[M/N]$ and in the study of the vector model in Section \ref{VM}.

\section{Higher order Padé Approximants for tree-level amplitudes}
\label{padessec}

\quad In this section we consider higher order Padé Approximants to
the partial wave amplitudes, with the hope that this will provide
some insight on the nature of the unitarization process discussed
above. We will see that the PA sequence associated with the IAM does
not converge properly, and that diagonal sequences are much more
suitable for this purpose. We begin with a brief overview of the
theory of PAs.

\subsection{Generalities on Padé Theory}

\quad Let $f(s)$ be a function with a Taylor expansion around $s=0$.
A Pade Approximant (PA) to $f(s)$, denoted as $P_N^M(s)$, is defined
as the ratio $P^M_N(s)=R_M(s)/Q_N(s)$ of two polynomials of orders
$M$ and $N$, respectively, and such that
$f(s)-P^M_N(s)=\cO(s^{M+N+1})$ when $s\to 0$.

The convergence properties of the PAs to a given function are more complex than those of the Taylor expansion. However, they converge for a broader
set of  functions, even in the case of slowly convergent or asymptotic power series, and they  usually carry smaller errors than the Taylor
expansions (when these are applicable). Furthermore,  in many cases, the Pad\'es  have been found to provide a fairly good approximation  even beyond
their expected range of applicability.

Pommerenke's theorem states that the sequence of diagonal PAs, i.e.
$[N/N]$, to a meromorphic function is convergent everywhere in any
compact set of the complex plane except, at most, for a zero capacity
set~\footnote{In addition to meromorphic, there are known theorems of convergence with
PAs for Stieltjes functions, continued functions, Gauss hypergeometric functions,
Bessel functions, some kind of divergent series, sets of complex
points, etc. This has been applied in the past to various kinds of scattering
processes~\cite{BakerEss}.}~\cite{Pommerenke,BakerEss,PerisPade06,Peris+Masjuan07,Peris+Masjuan08,BakerPades}.
This obviously includes the poles of $f(s)$, where the original
function is  ill-defined. In addition, the PA may produce a
series of poles absent  in $f(s)$. Thus, for a given compact region
${\mathcal K}$ in the complex plane, Pommerenke's theorem of
convergence requires that, either these undesired poles move away
from the region ${\mathcal K}$ as the order of the PA increases, or
they pair up with a close-by zero becoming what is usually called
a \textit{defect} \cite{BakerEss}. Although the PA breaks down in
the very neighborhood of these extraneous poles, away from them the
approximation is safe. Likewise, a PA can approximate a
multivaluated function (for example, a function with a logarithmic branch cut).
The sequence of PAs will  cluster its poles along the cut~\cite{BakerEss},
as we will see in the following section.

\subsection{Tree-level PAs in the LSM}

\quad In order to be able to handle the amplitude at higher orders,
we will consider the $\pi\pi$ scattering at tree-level. This is
equivalent to working in the limit $g \ll 1$ and keeping just the first
non-trivial contribution in the $g $ expansion. Thus, the
$\pi\pi$--scattering is determined in the LSM by the function
\begin{equation}
A(s,t,u)\,\, =
\,\, \Frac{s}{F^2}\,\Frac{M_\sigma^2}{M_\sigma^2\,-\,s}\, ,
\end{equation}
By means of the partial wave projection in Eq.~(\ref{eq.PW}), this provides
\begin{eqnarray}
t_0^0(s)  &=& \Frac{M_\sigma^2}{32 \pi F^2} \, \left[  - 5 + \Frac{ 3M_\sigma^2}{M_\sigma^2-s} + \Frac{ 2 M_\sigma^2}{s}
\ln\left(1+\Frac{s}{M_\sigma^2}\right)\right] \, , \nn\\[2mm]
t_1^1(s) &=&\Frac{M_\sigma^4}{32 s \pi F^2 } \, \left[  - 2 + \left(\Frac{ 2 M_\sigma^2}{s}+1\right)\, \ln\left(1+\Frac{s}{M_\sigma^2}\right)\right]
\, , \nn\\[2mm]
t_0^2(s) &=&- \Frac{M_\sigma^2}{16 \pi F^2 } \,
\left[ 1 - \Frac{  M_\sigma^2}{s}\, \ln\left(1+\Frac{s}{M_\sigma^2}\right)\right] \, .
\label{eq.LSM-PW}
\end{eqnarray}
The $\ln{\left[1+s/M_\sigma^2\right]}$ logarithms come from the partial-wave projection of the tree-level exchanges of resonances in the crossed
channel. They have absolutely nothing to do with the logarithms of the $\chi$PT amplitudes in Eq.~(\ref{eq.t4}), which come from the $\pi\pi$ loops.

At low energies the amplitude becomes
\begin{equation}
A(s,t,u)\,\, =\,\, \Frac{s}{F^2}\,\left[ 1\, +\, \Frac{ s}{M_\sigma^2} \, +\, \Frac{ s^2}{M_\sigma^4} \, \, +\,\, ...\right]\, ,
\end{equation}
so the partial waves are given by,
\begin{eqnarray}
t_0^0(s)  &=& \Frac{s}{16 \pi F^2} \, \left[ 1 + \Frac{11 s}{6 M_\sigma^2} +\Frac{15 s^2}{12 M_\sigma^4}\,\,+\,\,...\right] \, , \nn\\[2mm]
t_1^1(s) &=&\Frac{s}{96\pi F^2} \, \left[ 1- \Frac{s}{M_\sigma^2} +\Frac{ 9 s^2}{10 M_\sigma^4} \,\,+\,\, ... \right] \, , \nn\\[2mm]
t_0^2(s) &=&- \Frac{s}{32 \pi F^2}\, \left[ 1 - \Frac{ 2 s}{ 3 M_\sigma^2} + \Frac{s^2}{ 2 M_\sigma^4}\,\, +\,\,...\right] \, .
\end{eqnarray}
The comparison between the low-energy expansions and the whole result provides a first insight of the piece of information that is lost in the
unitarization procedure.  At high energies, the partial waves contain poles on the right-hand side of the $s$--plane, related to $s$--channel
resonance exchanges, and a left-hand cut, related to the crossed--channel resonance exchanges.
At low energies, both kinds of exchanges
contribute  equally to the low energy couplings, so the crossed resonance exchanges
shift the IAM poles from their  physical value.
Although $t$ and $u$ channels are not so relevant in the region close to the resonance pole, at low
energies they are  as important as the $s$--channel.

The simplest Pad\'e,  $P_1^1$, gives the prediction
\begin{eqnarray}
s_0^0 &=& \Frac{6}{11} M_\sigma^2\, ,\nn\\[2mm]
s_1^1 &=& - M_\sigma^2\, ,\nn\\[2mm]
s_0^2 &=& -\Frac{3}{2} M_\sigma^2 \, ,
\end{eqnarray}
which agrees with the one-loop calculation from Eqs.~(\ref{eq.LSM00})--(\ref{eq.LSM20}) if one remains at leading order in $g$.

\subsection{Higher order PAs in the LSM}

\quad The convergence of a sequence $[M/N]$ of Pad\'e Approximants
to a function implies that the PA tends to mimic its analytical
structure as $M,N\to \infty$. The common procedure for the
construction of a sequence of PAs is to increase $M,N$ following a
given pattern, e.g. $M=N\to\infty$. In some cases, this allows the application of
known mathematical theorems that ensure
convergence~\cite{Pommerenke,BakerEss,BakerPades}. This has been used for the study
of certain Green functions~\cite{PerisPade06,Peris+Masjuan07,Peris+Masjuan08}.
However,
little is known about the sequence $[M/N]$ when $M$ is kept fixed and $N\to\infty$ (for
instance $P^1_1,P^1_2...$).

%
%

We will employ and compare the $[1/N]$ and $[N/N]$ sequences for the study of
the $\pi\pi$ partial  wave scattering amplitudes. We will also comment
on PAs of the $[N+K/N]$, e.g  $[N-2/N]$.  In the next lines
we will focus our attention on the $IJ=00$ partial wave, but
analogous results are found for the other channels. Former
works pointed out that the PAs and other unitarizations fail to incorporate
the crossed channel resonance exchanges~\cite{PW,guo06}.
Nonetheless, we will see that as $N$ grows, the
poles of the sequence $[N/N]$ actually tend to mimic not only the
$s$--channel poles but also the left-hand cut contribution from
diagrams with resonances in the $t$ and $u$ channels.

\begin{figure}[!t]
\begin{center}
  \includegraphics[width=6cm]{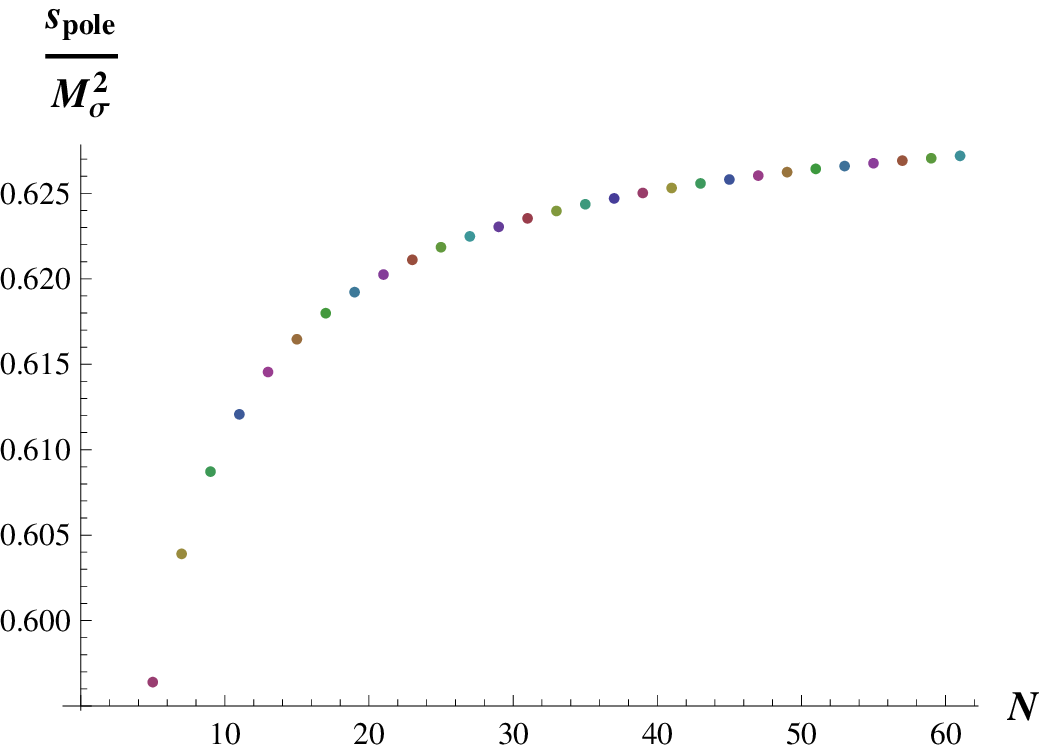}
  \hspace{2cm}
  \includegraphics[width=6cm]{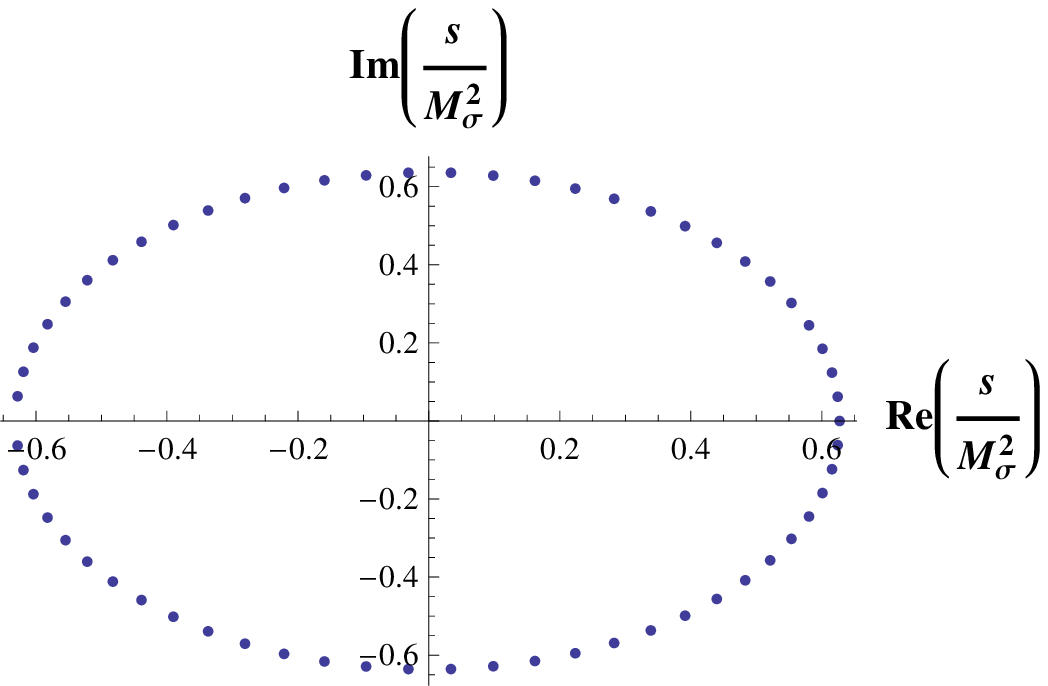}
  \caption{{\small   Left: Position of the nearest pole to $M_\sigma^2$
  for the first PAs of the form $[1/N]$ with $N$ odd
  (for even $N$ all the poles are complex).
  Right: Poles of the $P^1_{61}$ in the complex plane.}}
  \label{LSM30}
\end{center}
\end{figure}

The sequence $[K/N]$, with $K$ fixed, is studied in the present
section in the framework of the Linear Sigma Model. It provides  an
example of the behavior of these kind of sequences. $K=1$ is chosen
because of the similarity of this sequence and the
IAM~\cite{IAM,IAM-op6,IAM-P12-critics,IAM-poles,guo06}. Our results are
summarized in Fig.~\ref{LSM30}: No convergence is found with this
sequence.
In the case of $N$ odd, Fig.~\ref{LSM30}.a. shows that the $P^1_N$ pole
closest to $M_\sigma^2$ does not approach this value even for very
large $N$, always remaining a 30\% below. The analytical
structure of the original amplitude ($s$--channel sigma pole plus
left-hand cut) is never recovered since the $[1/N]$ PAs always set
the poles in the circular pattern  shown in
Fig.~\ref{LSM30}.b. This  suggests that
the use of further $[1/N]$ approximants to extend the IAM is not the optimal way to
proceed, even if  we had an accurate knowledge of the low-energy  expansion up to
very high orders.

\begin{figure}[!t]
\begin{center}
  \includegraphics[width=6cm]{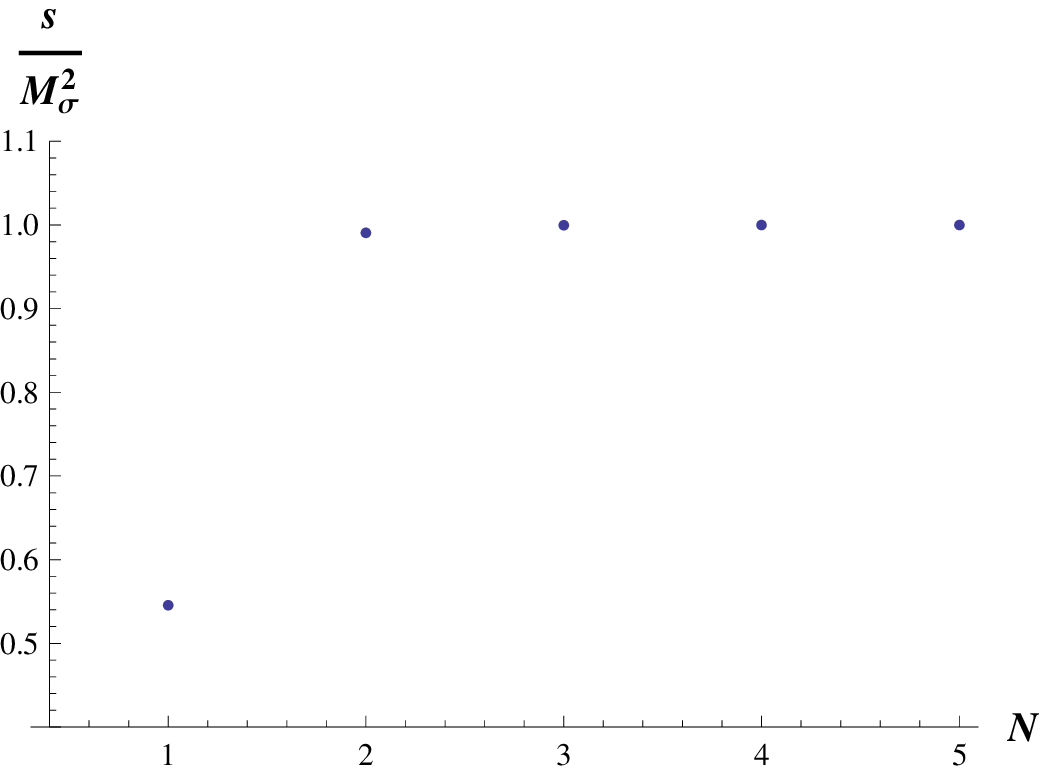}
  \hspace{1.5cm}
  \includegraphics[width=6cm]{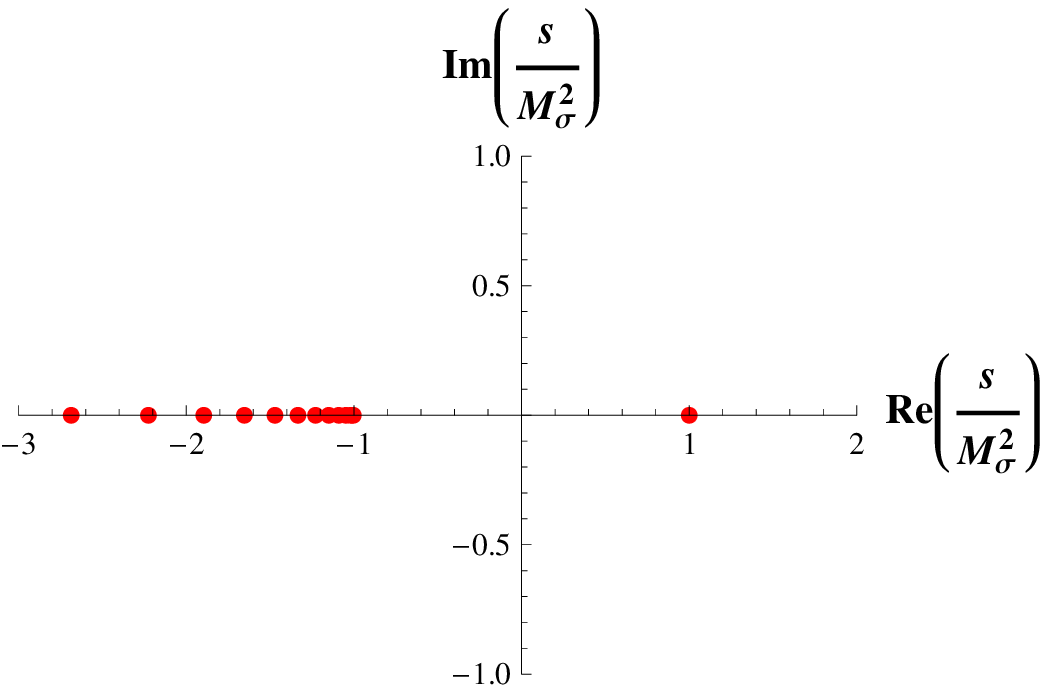}
  \caption{{\small
  Left: Location of the closest pole to $M_\sigma^2$
  for the first [N/N] PAs.
  Right: Poles of $P^{20}_{20}$.
  }}
\label{PadeNN20}
\end{center}
\end{figure}

Alternatively,  the use of sequences such as
$[N+K/N]$ (e.g. $[N-2/N],\, [N-1/N], \, [N/N],\,  [N+1/N]\ldots$)
seems to be a better strategy.
In the following we analyze the sequence $[N/N]$, as it ensures the
appropriate behavior  at high energies, $|t(s)|<1$. Nevertheless,
similar results have been generally found for the $[N+K/N]$ PAs with $K\neq 0$.
The $P^N_N$ pole closest to
$M_\sigma^2$ is shown in Fig.~\ref{PadeNN20}.a. One finds a quick
convergence of the sequence: $P_1^1$ reproduces the sigma pole a
$40\%$ off but $P^2_2$ disagrees by less than $1\%$, $P^3_3$ by less
than $0.1\%$, etc. Notice that already  $P_2^2$ provides a much
better description than $P^1_{61}$, although one includes far more
low-energy information in the latter. All this points out the
sizable discrepancy of the first element of the sequence ($P^1_1$)
with respect to the original amplitude. It also indicates that the $[1/N]$ PAs do
not produces a serious improvement. On the contrary, the $[N/N]$
sequence provide a far more efficient strategy with a quick
convergence.

Likewise,  Fig.~\ref{LSM30}.b shows how the $[1/N]$ PAs  are unable to recover
the analytical structure of the original amplitude, whereas the $[N/N]$ sequence,
besides providing the isolated pole of the sigma, tends to reproduce the left-hand cut as $N$ increases. The poles of $P_{20}^{20}$ are
plotted in Fig.~\ref{PadeNN20}. Although a PA is a rational function without cuts,
these are mimicked by placing poles where the cuts should lie.
The $P_{20}^{20}$ has one isolated pole near $M_{\sigma}^2$ (with an accuracy of $10^{-30}$) and nineteen poles over the real axis at
$s_p<-M_{\sigma}^2$, i.e. on the left-hand cut of the original function. As $N$ is increased, the number of poles lying on the branch cut increases
too.

A remarkable feature found for the first $P_N^N$ approximants ($P^1_1$, $P^2_2$, $P^3_3$)
is that they obey exact unitarity, as it happened with the IAM sequence $[1/N]$.

We would like to remark that,  although we lack of a general theorem
that ensures the convergence of the $[N+K/N]$ sequences, we have
found that they reproduce the original partial waves for arbitrary
$K$. Moreover, after performing modifications on the structure of
the amplitudes of Eq.~(\ref{eq.LSM-PW})  we still found convergence
the different $[N+K/N]$ PAs. In several situations the PAs set all
poles over the left-hand cut position and one isolated pole that
approached $M_\sigma^2$ when $N\to\infty$. In the worst cases, in
addition to this we found extraneous poles that either moved away as
$N$ increased or they tend to be canceled by nearby zeros at
$N\to\infty$.

As an amusement, we have also probed the PA sequence $[N/1]$ which
has the same number of inputs as the $[1/N]$ has for a given $N$.
In this new case we have found convergence in both LSM and the
resonance model presented in the following section but slower than
the $[N/N]$. For instance, the prediction for the $M_{\sigma}^2$ for
the first $P_1^N$ are
$\frac{s_p}{M_{\sigma}^2}=0.55,1.47,0.73,1.27,0.81...$ A criticism
that can be done to this sequence is its lack of unitarity, in
contrast to the other studied sequences.

\section{Vector Resonance Model}\label{VM}

\quad In order to broaden our analysis, we consider now a model with
just vector mesons~\cite{PW}. It could be derived either from the
gauged chiral model~\cite{Donoghue} for the couplings $3 g_\rho F^2=M_\rho^2$,  or
from resonance chiral theory~\cite{Ecker} with only vectors and the relation
$3 G_V^2=F^2$. The $\pi\pi$--scattering is given in this
model by
\begin{equation}
A(s,t,u) \, = \,  \Frac{M_\rho^2}{3 F^2}
\left[ \Frac{s-u}{M_\rho^2 -t} + \Frac{s-t}{M_\rho^2-u}  \right]\, .
\end{equation}

The study of the $IJ=11$ partial wave leads to the same conclusions found for the LSM.
It is at first sight remarkable that, on the contrary to the
previous case, one recovers $s_p=M_\rho^2$ from the first-order approximant $P_1^1$. However, the sequence $[1/N]$ already worsens at $N=2$, where
the two complex-conjugate poles are located at $s_p=(0.71\pm 0.96 i) \, M_\rho^2$ on the physical Riemann sheet. On the other hand, $P_N^N$ exactly
recovers $s_p=M_\rho^2$ for any odd $N$. For $N$ even, the prediction from $P^2_2$ is a 30\% off, but  one has again a quick convergence to
$s_p=M_\rho^2$ as $N$ increases: $P_4^4$ disagrees by less than 0.1\%, $P_6^6$ disagrees by less than $10^{-6}$, etc.

Furthermore, the $[1/N]$ and $[N/N]$ sequences produce, respectively, the same structure of poles found for the LSM. This is, $[1/N]$ generates the
circular structure of poles of Fig.~\ref{LSM30}.b.   and   the sequence $[N/N]$ places one pole at $s_p\simeq M_\rho^2$ and the remaining ones
reproducing the left-hand cut in analogy to Fig.~\ref{PadeNN20}.b.

\section{Discussion}\label{conc}

\quad In this letter, we have addressed the reliability of the
unitarization of low-energy amplitudes through Pad\'e Approximants.
The one-loop analysis of the LSM has led to IAM predictions of the sigma
mass and width respectively a factor 2 and 5 smaller than the
original ones in the model.

Some disagreement has been found in phenomenological determinations
of the $f_0(600)$ pole width~\cite{Leutwyler-criticism}, pointing
out that the IAM is unable to recover at the same time the value for
the mass and the width predicted from Roy equations~\cite{CCL}. This
discrepancy becomes even more obvious in the chiral limit, with
differences much larger than the expected $\cO(m_\pi^2)$
corrections.

It has been also shown that the $P^1_1$ mass prediction from the
one-loop analysis agrees at leading order with that coming from the
tree-level amplitude. This argument allowed  the construction of
higher order PAs based on the tree-level amplitude of the LSM and a
vector model. It was found that the sequence $[1/N]$ is unable to
reproduce the original partial wave scattering amplitude whereas the
PAs of the form $[N/N]$ display a quickly convergent behavior.
Moreover, though we lack of convergence theorem,
it was found that the $[N+K/N]$ PAs were able to reproduce
the partial waves for arbitrary $K$.

For all this, we suggest the use of the $[N/N]$ sequence rather than
$[1/N]$. Unfortunately, the study on broad resonances requires to go
beyond the tree-level approximation. Thus, in the real world, the
$IJ=00$ channel  needs the inclusion of loops, precluding by now the
extension to PAs beyond $\cO(p^4)$ in the chiral expansion, i.e.
$P^1_1$.  There is also a clear limitation on our experimental
knowledge of the low-energy couplings, which barely goes beyond  $\cO(p^4)$.
However, our proposal should be still suitable for the
analysis of theories with narrow resonances and a
relatively good knowledge of the experimental low-energy amplitudes.\\

\noindent {\bf Acknowledgements}\\

\noindent We would like to acknowledge helpful comments from S.~Peris, J.~Prades and  H.Q.~Zheng on the manuscript. We also want to thank A.~Fuhrer
and V.~Mateu for discussions concerning the LSM. This work has been partially supported by the EU-RTN Programme, Contract No. MRTN--CT-2006-035482
``Flavianet'', CYCIT-FEDER.FPA2005-02211, SGR2005-00916 and by the Spanish Consolider-Ingenio 2010 Program CPAN (CSD2007-00042).

\end{document}